\documentstyle[aps,prl]{revtex}

\begin{document}

\draft

\title{Condensate fraction and critical temperature 
of a trapped interacting Bose gas}

\author{ S. Giorgini$^{1}$, L. P. Pitaevskii$^{2,3}$ and S. Stringari$^{1}$}

\address{$^{1}$Dipartimento di Fisica, Universit\`a di Trento, \protect\\
and Istituto Nazionale di Fisica della Materia, I-38050 Povo, Italy}
\address{$^{2}$Department of Physics, Technion, 32000 Haifa, Israel}
\address{$^{3}$Kapitza Institute for Physical Problems, 117454 Moscow, Russia}


\maketitle

\begin{abstract}

{\it By using a mean field approach, based on  the Popov approximation,
we calculate the temperature dependence of the condensate fraction 
of an interacting Bose gas confined in an anisotropic 
harmonic trap. For systems 
interacting with repulsive forces we find
a significant decrease of the condensate fraction and of the critical 
temperature with respect to the predictions of the non-interacting model. 
These  effects go in the opposite direction 
compared to the case of a homogeneous gas.   
An analytic result
for the shift of the critical temperature holding to first order in the
scattering length is also derived.}

\end{abstract}


\narrowtext

The recent experiments on Bose-Einstein condensation (BEC) in magnetically 
trapped atomic vapours \cite{EXP}
have stimulated a new interest in the 
theoretical study of inhomogeneous Bose gases. Although the atom clouds 
realized in these experiments are very dilute, 
the effects due to the interatomic forces 
are known to be important at low temperature.
In particular, the shape and the energy of the condensate cloud 
\cite{BUR,DS} as well as the dispersion law of the 
elementary excitations
\cite{CEX} are strongly affected by the interaction. 
In the very recent experiments by the Boulder and MIT groups  
\cite{KET0,COR,KET}, the measured 
release energy and  excitation 
frequencies of the collective modes have been found to be in good agreement
with the theoretical predictions, thereby revealing 
important features of the trapped Bose condensed gases which are
undoubtedly connected to the interparticle interaction.
The question of how two-body forces affect the thermodynamic
properties of these systems has been also the object of 
several theoretical investigations \cite{FT}.
The critical temperature 
of the BEC transition in an homogeneous dilute gas 
has been recently calculated \cite{BS} using
the renormalization group theory. The result is a shift towards higher 
temperatures, with respect to the prediction of the ideal Bose gas.
Similar results have been also obtained with 
path integral Monte Carlo simulations \cite{GRU}.
However, 
no definitive conclusions have been so far drawn on the behavior
of the condensate fraction, nor of the critical temperature
in the presence of a confining potential \cite{KRA}. First experimental data 
on these
relevant quantities are now becoming available \cite{KET0,BOU}.

Finite size effects on the temperature 
dependence of the condensate fraction and on the critical temperature 
in the presence of an external trap 
have been recently investigated by several authors within the 
non-interacting model \cite{KT,IG}.
In the presence of an anisotropic harmonic potential of the form $V_{ext}=m
(\omega_x^2 x^2+\omega_y^2 y^2+\omega_z^2 z^2)/2$ this model predicts, in
the large $N$ limit, the well known results for the critical temperature 
\begin{equation}
T_c^0 = \frac{\hbar\omega}{k_B}\; 
(\frac{N}{\zeta(3)})^{1/3} \simeq 0.94 \frac{\hbar\omega}{k_B}\;N^{1/3}
\;\;,
\label{Tc0}
\end{equation}
where $\omega=(\omega_x\omega_y\omega_z)^{1/3}$ is the geometrical average
of the oscillator frequencies, and for the temperature
dependence of the number of atoms in the condensate
\begin{equation}
\frac{N_0(T)}{N} = 1 - \left(\frac{T}{T_c^0}\right)^3  \;\;.
\label{cf0}
\end{equation}
Results (\ref{Tc0}) and (\ref{cf0}) are obtained using the
semiclassical approximation for the excited states and setting the chemical 
potential equal to zero at the transition.
The first correction to the critical temperature (\ref{Tc0}) due to the finite 
number of atoms in 
the trap 
has been recently shown \cite{IG} to obey the law
\begin{equation}
\frac{\delta T_c^0}{T_c^0} = - \frac{\zeta(2)}{2\zeta(3)^{2/3}}\;
\frac{\bar{\omega}}{\omega} \;
N^{-1/3} \simeq - 0.73 \; \frac{\bar{\omega}}{\omega} \;
N^{-1/3} \;\;,
\label{idgas}
\end{equation}
where $\bar{\omega}=(\omega_x+\omega_y+\omega_z)/3$ is the mean frequency. 
This result can be obtained by still employing the semiclassical description 
for the excited states, but keeping the quantum value $\mu = 3\hbar\bar{\omega}
/2$ for the chemical potential at the transition. The discretization of the
excited energy levels gives rise to higher order corrections to 
$\delta T_c^0$. 

In this Letter we present results for the temperature dependence of the 
condensate fraction and for the critical temperature of a dilute Bose gas 
interacting with repulsive forces and confined in an 
harmonic potential. We use a mean field approach and 
the semiclassical approximation for the excited states. 
This approximation is expected to be accurate for temperatures significantly 
larger than the oscillator temperature $\hbar\omega/k_B$. This condition is
well satisfied in a useful range of temperatures provided $N$ is sufficiently
large.
In the presence of repulsive interactions we find that the thermal
depletion of the condensate is enhanced and 
the critical temperature is shifted towards lower temperatures. These results 
go in the opposite direction compared to the homogeneous case, revealing 
an interesting behavior exhibited by a confined Bose gas.

Our starting point is the finite temperature 
generalization of the Gross-Pitaevskii
equation within the Popov approximation \cite{POP,GRI}
\begin{eqnarray}
- \frac{\hbar^2\nabla^2}{2m}\Phi({\bf r}) 
&+& (\; V_{ext}({\bf r}) - \mu 
\nonumber \\
&+& g(n_0({\bf r}) + 2n_T({\bf r}))
\; )\Phi({\bf r}) = 0 \;\;, 
\label{GP}
\end{eqnarray}
and  
\begin{eqnarray}
i\hbar\frac{\partial \Phi'({\bf r},t)}{\partial t} &=& - 
\frac{\hbar^2\nabla^2}
{2m}\Phi'({\bf r},t) \;+\; ( \;V_{ext}({\bf r}) -\mu 
\nonumber \\
&+& 2gn({\bf r})  
\;)\Phi'({\bf r},t) 
\;+\; gn_0({\bf r})\Phi'^*({\bf r},t) \;\;.
\label{eex}
\end{eqnarray}
The first equation describes the space variations of the condensate wave 
function 
$\Phi({\bf r})
= \langle \psi({\bf r}) \rangle$ at statistical equilibrium, 
where $\psi({\bf r})$ is the particle field operator.
The second one is the equation for the fluctuations 
of the condensate $\Phi'({\bf r},t) = <\psi({\bf r},t)> - \Phi({\bf r})$ which 
give
the elementary excitations of the system.
In Eqs. (\ref{GP}) and (\ref{eex}), $V_{ext}$ is the external potential, 
$\mu$ is the chemical potential, $g=4\pi\hbar^2a/m$ is the 
interaction coupling constant fixed by 
the $s$-wave scattering length $a$, $n_0$ is the equilibrium condensate 
density $n_0({\bf r})=|\Phi({\bf r})|^2$,  
$n({\bf r})=\langle\psi^{\dagger}({\bf r})
\psi({\bf r})
\rangle$ is the particle density and finally $n_T$ is the density of the 
thermally excited particles $n_T({\bf r}) = n({\bf r}) - n_0({\bf r})$. 
Equations (\ref{GP}), (\ref{eex}) are obtained from 
the equation of
motion for the particle field operator $\psi({\bf r},t)$, by treating the cubic
interaction term $g\psi^{\dagger}({\bf r},t)\psi({\bf r},t)\psi({\bf r},t)$ 
in the mean field scheme. 
The Popov approximation
consists in neglecting the anomalous
density $m_T({\bf r})=\langle\psi({\bf r})\psi({\bf r})\rangle - \Phi({\bf r})
^2$ entering 
the interaction term. 
As discussed in Ref.
\cite{GRI} this approximation is expected to be good for high temperatures
where $m_T<<n_T$ and also in the low-temperature
regime where $n_T$ and $m_T$ are of the same order, but both are negligible. 
The present mean field approach is expected to provide correctly
the thermodynamic properties of the system, apart from the critical behavior
near $T_c$ where the mean field approach is known to fail.

The energies of the elementary excitations can be explicitly 
obtained from Eq. (\ref{eex}) using the semiclassical
WKB approximation. 
Let us write the fluctuations of the condensate in the form
$\Phi'({\bf r},t) = A \exp(i\varphi({\bf r},t)) + B \exp(-i\varphi({\bf r},t)) 
$.
In WKB approximation 
the coupled equations for $\Phi'({\bf r},t)$ 
and 
$\Phi'^*({\bf r},t)$, given by Eq. (\ref{eex}) and its complex conjugate, can 
be 
solved in an explicit way, yielding the semiclassical  
excitation spectrum 
\begin{eqnarray}
&\epsilon({\bf p},{\bf r})&
\nonumber \\
&=& \sqrt{\left(\frac{p^2}{2m} + 
V_{ext}({\bf r}) - 
\mu + 2g
n({\bf r}) \right)^2  
- g^2n_0^2({\bf r})}  \;,
\label{bge}
\end{eqnarray}
where $p=\hbar\nabla\varphi$ and
$\epsilon=-\hbar\partial\varphi/\partial t$ 
are respectively the impulse and the 
energy
of the excitation. For a homogeneous system at low temperatures, where
$n_T$ can be neglected, the above excitations coincide with the usual 
Bogoliubov spectrum. 
The quasi-particles with energies $\epsilon({\bf p},{\bf r})$ are distributed 
according
to the Bose distribution function
\begin{equation}
f({\bf p},{\bf r}) = \frac{1}{\exp(\epsilon({\bf p},{\bf r})/k_BT) - 1} \;\;,
\label{qpf}
\end{equation}
whereas the particle distribution function can be obtained from the Bogoliubov
canonical transformations and is given by 
\begin{equation}
F({\bf p},{\bf r}) = - \left( \frac{\partial \epsilon}{\partial \mu} 
\right)_{n_0} f({\bf p},{\bf r}) 
\;\; ,
\label{pf}
\end{equation}
where the partial derivative is taken at constant condensate density $n_0$.
The non-condensate density is then obtained by integrating $F({\bf p},{\bf r})$ 
in momentum space
\begin{equation}
n_T({\bf r}) = \int \; \frac {d^3{\bf p}}{(2\pi\hbar)^3} \; 
F({\bf p},{\bf r}) \;\;, 
\label{nt}
\end{equation}
and the total number of atoms out of the condensate is given by
\begin{equation}
N_T = \int \; d^3{\bf r} \; n_T({\bf r}) \;\;.
\label{Nt}
\end{equation} 
In the above equations the 
chemical potential $\mu$ is fixed by the normalization condition
\begin{equation}
N = N_0(T) + N_T \;\;,
\label{norm}
\end{equation}
where $N_0(T) = \int\;d^3{\bf r} \; n_0({\bf r})$ is the number of atoms in the 
condensate, with $n_0({\bf r})$ fixed by the solution of Eq. (\ref{GP}).

We have solved Eqs. (\ref{GP}) - (\ref{norm}) in a self-consistent way
employing the following procedure: \\
i) Equation (\ref{GP}) is solved, using the method described 
in Ref. \cite{DS}, for the
condensate density $n_0({\bf r})$ and the chemical potential $\mu$, by keeping
fixed the number $N_0(T)$ of particles in the condensate and the density 
$n_T({\bf r})$ of thermally excited atoms. \\
ii) The condensate density and chemical potential found in step 1) are used to
calculate the excitation energies from Eq. (\ref{bge}). A new 
density $n_T$ is then obtained from Eq. (\ref{nt}), and  new values for the 
number of atoms out of the condensate, $N_T$, and in the condensate, $N_0(T)$,
are derived from Eqs. (\ref{Nt}) and
(\ref{norm}). \\ 
iii) Step 1) and step 2) are repeated until convergence is reached.

It is worth noticing that the present method accounts for finite size effects 
because of the quantum 
nature of Eq. (\ref{GP}) for the order parameter and of the corresponding 
value of the chemical potential. In particular, in the absence of interactions,
it reproduces result (\ref{idgas}) for the shift of the 
critical temperature.

In Fig. 1 we present our results for the temperature dependence of the 
condensate fraction $N_0(T)/N$ for a system of 5000 Rb atoms interacting with 
scattering 
length  
$a=110 a_0$, where 
$a_0$ is the Bohr radius, and trapped in an axially deformed harmonic 
potential fixed by   
$a_{HO} = (\hbar/m\omega)^{1/2} = 7.92 \times 10^{-5}\;$ cm and by the 
deformation parameter $\lambda=\omega_z/\omega_x=\omega_z/\omega_y = 
\sqrt{8}$.
These values correspond to the experimental conditions of Ref. \cite{COR}.  
As clearly emerges from the figure, finite size effects are not 
negligible for this value of $N$. However, interaction effects are more 
important and result in a sizable  quenching of the condensate fraction.
Our results can be understood by looking at the role of the 
two-body interaction in Eqs. (\ref{bge}) - (\ref{Nt}). The chemical potential 
is significantly 
enhanced, thereby favouring the thermal depletion; on the other hand the 
interaction terms in $g$ of Eq. (\ref{bge}) have a minor effect because of the 
small 
overlap between $n_0({\bf r})$ and $n_T({\bf r})$. This behaviour differs from
the one exhibited by a homogeneous dilute Bose gas where one finds a reduction 
of the thermal depletion with respect to the prediction of the non-interacting
model \cite{BS,GRU}.

When the number of atoms in the trap increases, the effects due to the 
interaction become more and more important,
as explicitly shown in Fig. 2, where we report results
for $N_0(T)$ corresponding to three values of $N$ ($N=10^5$, $10^6$, $10^7$).
The calculations of Fig. 2 have been carried out using the value $2.56 \times
10^{-3}$ for the ratio $a/a_{HO}$ and $\lambda=18/320$. This choice 
corresponds to the experimental situation of 
Ref. \cite{KET0} for Na atoms.
We have checked that for such large values of $N$ the results
for  $N_0(T)$ are practically independent  
of the value of the deformation parameter $\lambda$. 
Notice that the quenching effects shown in Fig. 2 are quantitatively similar 
to the ones of Fig. 1, despite the much larger values of $N$ contained in the
trap. 
This is due to the fact that in the trap of Fig. 2 (MIT-type trap) the ratio
$a/a_{H0}$ is a factor $3$ smaller than in the trap  of Fig. 1 (JILA-type trap).
 
The shift $\delta T_c$ in the critical temperature can  
be estimated analytically to the lowest order
in the coupling constant $g$ by studying the behavior of the trapped gas  
for temperatures $T \geq T_c$. In this case 
the order parameter $\Phi({\bf r})$ vanishes and the dispersion
of the  elementary 
excitations takes the simple Hartree-Fock form 
\begin{equation}
\epsilon({\bf p},{\bf r}) = \frac {p^2}{2m} + V_{ext}({\bf r}) + 2gn_T({\bf r})
- \mu \;\;.
\label{eext}
\end{equation}
To the lowest order in $g$  
the total number of particles above $T_c$ 
can be written in the following form 
\begin{equation}
N = N_T^0(\mu) - 2g \int \;d^3{\bf r} \;\frac{\partial n_T^0}
{\partial \mu} n_T^0({\bf r},\mu) \;\;,
\label{shift1}
\end{equation}
where we have used Eqs. (\ref{qpf}) - (\ref{Nt}) with 
$\epsilon({\bf p},{\bf r})$ given 
by Eq. (\ref{eext}).
The quantity $n_T^0({\bf r},\mu)$ is the non-condensed density 
given by the non-interacting model
\begin{equation}
n_T^0({\bf r},\mu) = \lambda_T^{-3} g_{3/2}(\exp(-(V_{ext}({\bf r})-\mu)/k_BT)) 
\;\;,
\label{dig}
\end{equation}
where $\lambda_T=\hbar(2\pi/mk_BT)^{1/2}$ is the thermal wavelength and
$g_{3/2}(x) = \sum_{n=1}^{\infty} x^n/n^{3/2}$.  

Bose-Einstein condensation starts at the temperature $T_c$ for which the 
chemical potential reaches 
the energy of the lowest solution of the 
Schr\"odinger equation corresponding to the Hamiltonian (\ref{eext}).  
To the lowest order in $g$ and for large values of $N$ one finds
\begin{equation} 
\mu=\frac{3\hbar\bar{\omega}}{2} + 2gn_T^0({\bf r}=0,\mu=0) \;\;.
\label{mu}
\end{equation}
By expanding the r.h.s. of Eq. (\ref{shift1}) around $\mu=0$ and $T=T_c^0$, 
we obtain the following 
relationship for the shift $\delta T_c=T_c-T_c^0$ of the critical
temperature 
\begin{eqnarray}
\delta T_c  \int \; d^3{\bf r} \frac{\partial n_T^0}{\partial
 T} =
- \frac{3\hbar\bar{\omega}}{2} \int \; d^3{\bf r} \; \frac{\partial n_T^0}
{\partial\mu} \;\;\;\;\;\;\;
\nonumber \\
\;\;\;\;\;\;\;
- 2g \int \; d^3{\bf r} \;\frac{\partial n_T^0}{\partial \mu} \;(n_T^0({\bf r}
=0) -
n_T^0({\bf r})) \;\;,
\label{shift2}
\end{eqnarray}
where, for simplicity,  we have dropped the argument $\mu=0$ from the  
$n_T^0$ functions, here evaluated at $T=T_c^0$. 
The shift $\delta T_c$ of the critical temperature is hence the sum of two 
distinct effects 
given by the two terms on the r.h.s. of Eq. (\ref{shift2}). The first 
contribution is due to the finite number of particles 
present in the trap and after a straightforward integration gives exactly 
result
(\ref{idgas}) for the shift $\delta T_c^0/T_c^0$.  
The second contribution instead arises from interaction effects and takes the 
form
\begin{eqnarray}
\frac {\delta T_c^{int}}{T_c^0} &=&  
- \frac{2g}{T_c^0} \; \frac{\int\;d^3{\bf r}\;\partial n_T^0 / \partial\mu
\;(n_T^0({\bf r}=0) - n_T^0({\bf r}))} {\int\;d^3{\bf r}\;
\partial n_T^0 / \partial T} \nonumber \\ \; &=& - 
\frac{a}{a_{HO}}\;N^{1/6}\;
\frac {2^{3/2}\zeta(3/2)\zeta(2)}{3\sqrt{\pi}\zeta(3)^{7/6}} (1 - S) \;\;,
\label{shift3}
\end{eqnarray}
where $S = \sum_{n,m=1}^{\infty} \frac{1}{n^{1/2}m^{3/2}} 
\frac{1}{(n+m)^{3/2}}/ \zeta(3/2)\zeta(2) \simeq 0.281$.
By evaluating explicitly the numerical factors one gets the relevant result
\begin{equation}
\frac {\delta T_c^{int}}{T_c^0}  
\simeq - 1.33 \;\frac{a}{a_{HO}}\;N^{1/6}
\;\;.
\label{shift4}
\end{equation}
The shift $\delta T_c^{0}$, originating from finite size effects, is always 
negative and vanishes in the large $N$
limit. For an axially deformed trap it depends on the deformation parameter 
$\lambda = \omega_z/\omega_x = \omega_z/\omega_y$
through the ratio $\bar{\omega}/\omega = (\lambda+2)/3\lambda^{1/3}$ and is 
minimum for an isotropic trap. 
Vice versa the shift $\delta T_c^{int}$ due to interactions  
can be either 
negative
or positive, depending on the sign of $a$, it increases 
as $N^{1/6}$ and does not depend on the deformation parameter $\lambda$, 
but only on the geometrical average $\omega$. 
Furthermore it  
vanishes   
for a homogeneous system (see Eq. (\ref{shift3}) ) where the density $n_T^0$ is 
uniform, thereby showing that a shift in the critical temperature, linear
in the interaction coupling constant, can occur only in the
presence of an external trap.
It is also worth noticing that in the present approach the relationship between
the critical temperature and the corresponding value of the density in the 
center of the trap is unaffected by the interaction. The shift (\ref{shift4})
is hence always associated with a change in the central density produced 
by interparticle forces.  
The prediction $\delta T_c = \delta T_c^{0} + 
\delta T_c^{int}$, obtained from Eqs. (\ref{idgas}) and (\ref{shift4}), 
well agrees with
the numerical values 
obtained in the self-consistent calculation discussed
above and reported in Figs. 1 and 2. 
The relative importance of the two effects depends on the value of $N$. 
Notice however that the interaction effects depend very weakly on $N$ and scale
as $N^{1/6}a/a_{HO}$. This behaviour should be compared with the  
effects of the interactions 
on the low-temperature properties of the system (e.g. the size and the
energy of the condensate) which instead scale \cite{DS} as $Na/a_{HO}$.

We finally note that according to the Ginzburg criterion critical 
fluctuations violate the mean field result in the region 
\begin{equation}
\delta\mu = |\mu - \mu_c| \leq \frac {m^3 g^2 {T_c^0}^2}{\hbar^6} \;\; 
\label{gin}
\end{equation}
where $\mu_c$ is the value of the chemical potential at the critical point.
These fluctuations can affect the shift in the chemical potential at the 
critical point only to order $g^2$. It means that the 
mean field approach can provide a 
reliable 
prediction of the shift of the critical temperature up to terms linear in the 
scattering length.

A more systematic discussion of the thermodynamic behavior of a trapped Bose
gas,
including the temperature dependence of the density profiles, 
the specific heat and  
the superfluid density will be the object of a future paper.

Useful discussions with Franco Dalfovo are acknowledged.

\begin{figure}
\caption{Temperature dependence of the condensate fraction for a system of 5000
Rb atoms. The solid line is the result for the interacting case, the 
short-dashed line is the exact result for the non-interacting case, 
and the long-dashed
line corresponds to the non-interacting case in the thermodynamic limit 
(Eq. (2)).}
\end{figure}

\begin{figure}
\caption{Temperature dependence of the condensate fraction for interacting 
Na atoms. The solid line corresponds to $N=10^7$, the short-dashed line 
corresponds to $N=10^6$ and the long-dashed line to $N=10^5$. The dot-dashed
line is the result for the non-interacting case in the thermodynamic limit 
(Eq. (2)).}
\end{figure}

\end{document}